\documentclass[letterpaper]{jpconf}
\usepackage{graphicx}
\begin{document}
\title{Parameter-free Stark Broadening of Hydrogen Lines in DA White Dwarfs}

\author{P-E Tremblay$^1$, P Bergeron$^1$ and J Dupuis$^2$}

\address{$^1$ D\'epartement de Physique, Universit\'e de Montr\'eal, C.P.~6128, Succ.~Centre-Ville, Montr\'eal, Qu\'ebec H3C 3J7, Canada}
\address{$^2$ Canadian Space Agency, 6767 Route de l'A\'eroport, Longueuil, Qu\'ebec J3Y 8Y9, Canada}

\ead{tremblay@astro.umontreal.ca, bergeron@astro.umontreal.ca, jean.dupuis@space.gc.ca}

\begin{abstract}
We present new calculations for the Stark broadening of the hydrogen
line profiles in the dense atmospheres of white dwarf stars. Our
improved model is based on the unified theory of Stark broadening from
Vidal, Cooper \& Smith, but it also includes non-ideal gas effects from
the Hummer \& Mihalas occupation probability formalism directly inside
the line profile calculations. This approach improves upon previous
calculations that relied on the use of an ad-hoc free parameter to
describe the dissolution of the line wing opacity in the presence of
high electric microfields in the plasma. We present here the first
grid of model spectra for hot ($T_{\rm eff} \ge 12,000$~K) DA white
dwarfs that has no free parameters. The atmospheric parameters
obtained from optical and UV spectroscopic observations using these
improved models are shown to differ substantially from those published
in previous studies.
\end{abstract}

\section{Introduction}

The spectroscopic technique has been the most successful method for
determining the atmospheric parameters --- $T_{\rm eff}$ and $\log g$
--- of hydrogen-line DA white dwarfs \cite{bergeron92}. The technique
consists in comparing the observed and predicted Balmer line profiles,
and it has recently been expanded to the Lyman line series in the
ultraviolet \cite{hebrard03}. The advantage of
this method is that the theoretical profiles are very sensitive to
variations of the atmospheric parameters. Since about 80\% of the
white dwarf population is of the DA type, the spectroscopic technique
coupled with high signal-to-noise spectroscopic observations of Lyman
or Balmer lines for large samples of DA stars can reveal important
details about the mass distribution, the luminosity function, and the
formation rate of white dwarfs. Even though this technique yields very
accurate results with uncertainties less than 2$\%$ in terms of the
{\it relative} parameters between stars, the {\it absolute} values of
these parameters may still depend on the physics included in the
models. One important aspect of the model atmosphere calculations is
the opacity of the hydrogen lines, which are the dominant features
observed in optical and UV spectra.

\section{Models with a free parameter}

In their preliminary analysis of DA white dwarfs,
\cite{bergeron92} found a lack of internal consistency between
the spectroscopic solutions obtained when an increasing number of
Balmer lines was included in the fitting procedure. This is
illustrated in the top panel of Figure 1 for a typical DA star where
we can see the solution gradually drifting in the $T_{\rm eff}-\log g$
diagram as more lines are included in the fit. \cite{bergeron93}
traced back the problem to the neglect of non-ideal gas effects inside
the line profile calculations. Indeed, all white dwarf models rely on the
theory of Stark broadening from Vidal, Cooper, \& Smith \cite{vcs} (hereafter
VCS), which assumes that the absorber is in an ideal gas. However,
\cite{seaton90} has argued that non-ideal effects resulting from
the perturbations of neighboring particles should be included {\it
directly} in the line profile calculations. As a simpler
alternative, \cite{bergeron93} proposed instead to parameterize the
value of the critical field ($\beta_{\rm crit}$) used in the non-ideal
equation of state of Hummer \& Mihalas
\cite{hm88,dappen87} (hereafter HM88) to {\it mimic} the non-ideal
effects in the line opacity, and in particular in the regions where
the line wings overlap.  Model spectra calculated with twice the value
of the critical field ($\beta_{\rm crit}\times2$) were shown to
improve the internal consistency significantly, as can be observed in the
middle panel of Figure 1. It should be stressed, however, that the
use of this free parameter does not imply that HM88 have
underestimated the value of the critical field.  As explained in detail by
\cite{bergeron93}, this is just a way to simulate the non-ideal
effects discussed in \cite{seaton90} by {\it reducing} the line wing
opacity near the Balmer limit. The effect of this additional free
parameter on the line wing opacity is illustrated in Figure 2 where
two model spectra are compared. The problem with this approach, of
course, is that it has no physical basis, and there is also no reason
to believe that this free parameter is the same for various hydrogen
states or for different atmospheric parameters. It also has the
disadvantage of modifying artificially the various populations in the
equation of state. We note that all currently available model spectra
of DA white dwarfs
\cite{bergeron91,hubeny95,finley97,vennes05} make use of this free parameter.
Finally, even though the same
parameterization is also used for transitions in the UV, it
has never been tested 
empirically in an analysis similar to that
displayed in Figure 1.

\begin{figure}[h]
\begin{minipage}{18pc}
\includegraphics[width=19pc]{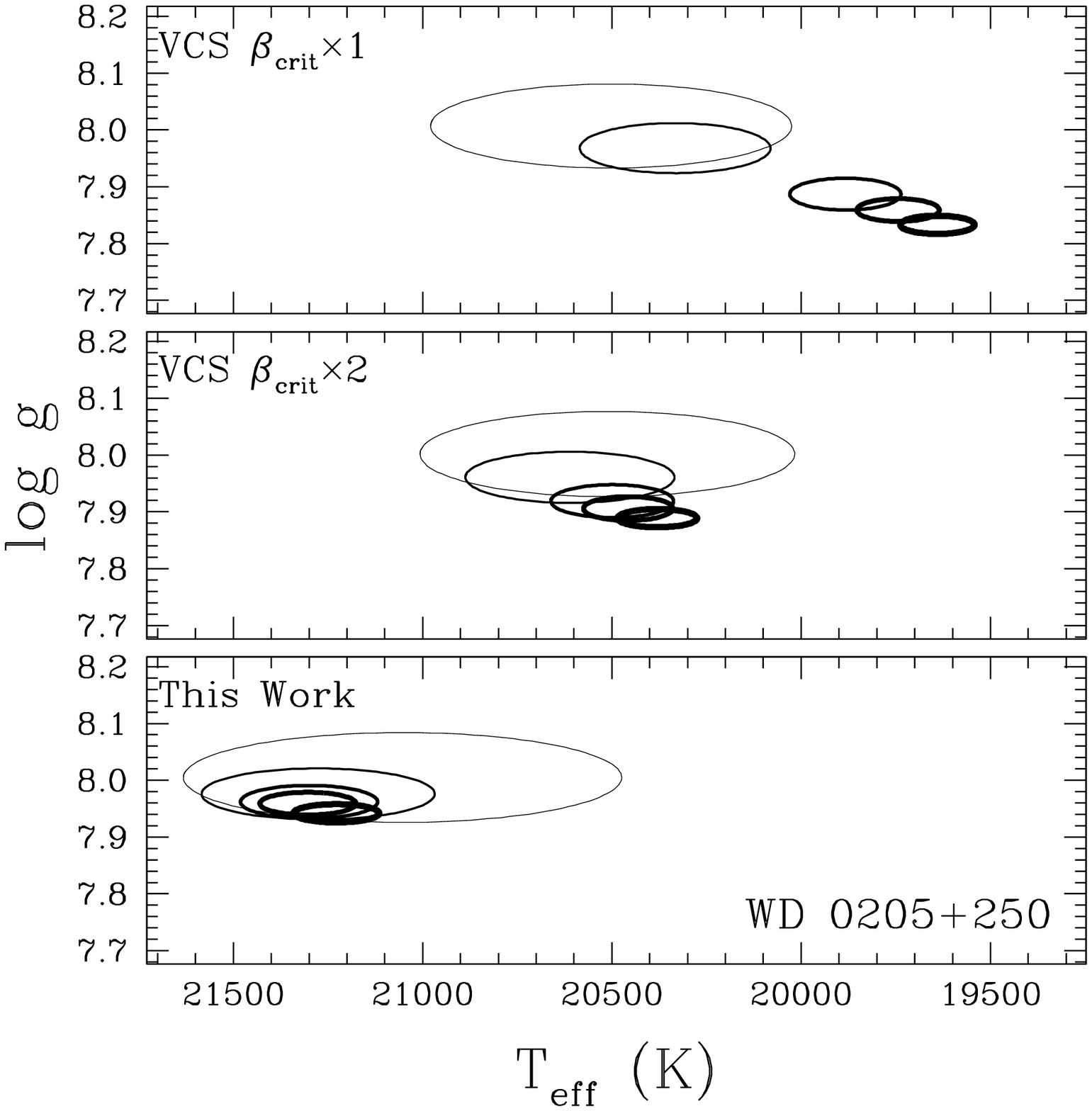}
\caption{Solutions in a $T_{\rm eff}-\log g$ diagram for a typical
DA star using 1 line (H$\beta$), 2, 3, 4 and 5 lines (up
to H8) in the fitting procedure (represented by thicker
1$\sigma$ uncertainty ellipses). The top panel shows the results with the original
VCS profiles, while the middle panel also includes the free parameter proposed
by \cite{bergeron93}. The bottom panel is with our new line profiles
(see \S~3).}
\end{minipage}\hspace{2pc}
\begin{minipage}{19pc}
\includegraphics[width=18pc]{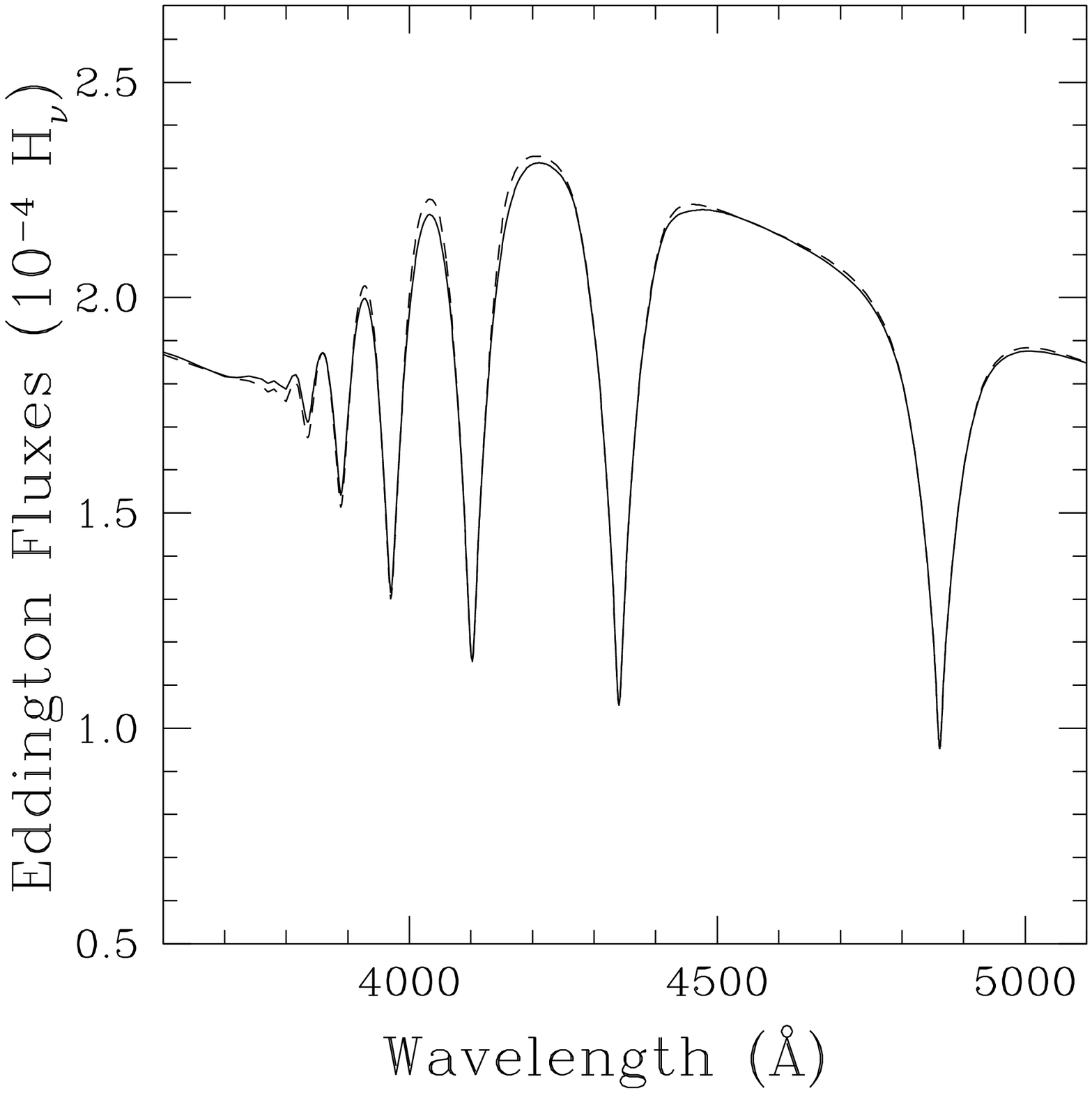}
\caption{Comparison of synthetic spectra at $T_{\rm eff}$ = 20,000 K and 
$\log g$ = 8 without (solid line) and with (dashed line) the
parameterization of the critical field ($\beta_{\rm crit}\times2$)
proposed by \cite{bergeron93}.}
\end{minipage} 
\end{figure}

\section{Improved Stark profiles}

In this work, we provide a framework with no free parameter by
improving the physics in the line profile calculations. We briefly
describe here our procedure; further details will
be provided elsewhere. Our work is based on the unified theory of
Stark broadening from VCS. An extension of the VCS calculations in the
white dwarf regime has been published by
\cite{lemke97}, and these tables have been the most commonly used to model the
Balmer and Lyman line profiles in DA stars. The unified VCS theory
relies on the quasi-static approximation to describe the broadening by
protons and a unified classical path theory for the broadening by
electrons, which reduces to the impact approximation in the line cores
and to the quasi-static approximation in the wings. One major problem
for the modeling of DA atmospheres, however, is that the theory
considers the absorber to be in an ideal gas, which means that line
dissolution due to plasma perturbations is not allowed. Such
perturbations are described at length in the HM88 non-ideal equation of state
for astrophysical applications. They include, in particular, the
high amplitude electric microfields produced by protons and the frequent
electronic collisions.

The occupation probability formalism of HM88 has been introduced in
the context of white dwarf spectra since the work of
\cite{bergeron91}. According to this formalism, the probability that
one electronic state is bound to the atom is obtained from the
integral of the electric microfield probability distribution up to a
given value for the critical field $\beta_{\rm crit}$. The critical
field is roughly defined as the point where one Stark state for one
particular level with a principal quantum number $n$ ``crosses''
another Stark state associated with a higher level $n+1$. At the
critical field, we expect that a bound electron on a level $n$, and a
superposition of the accessible Stark states due to the electronic
collisions and fluctuations in the direction of the microfield, will
undergo a cascade of transitions to higher levels all the way up to
the continuum. The occupation probability obtained with this formalism
can then be used to compute the hydrogen state populations. The
treatment of the amplitude of the line opacity must also
be modified to take into account the fact that a transition to an
upper ``dissolved'' atomic level must be treated as a bound-free
opacity rather than a bound-bound transition --- the so-called
pseudo-continuum opacity
\cite{dappen87}.

The late Mike Seaton was the first to introduce the HM88 non-ideal
effects directly inside the line profile calculations so that both the
spectral shape and the amplitude are affected in a consistent way
\cite{seaton90}. However, his calculations made for the Opacity
Project rely on an approximate electronic broadening theory that is
inappropriate in the context of white dwarf atmospheres. We have
therefore performed our own implementation of these non-ideal effects
in a version of the VCS code provided to us by M. Lemke. In the case
of proton perturbations, the line wing opacity will be reduced with
respect to the ideal gas case since the high electric microfields
contribute more importantly to the bound-free opacity by dissociating
the atoms. We therefore use here a truncation of the proton
microfields in the calculations of the quasi-static proton line
broadening and include corrections due to electron perturbations as
well. This is the first time these non-ideal effects are included
directly into the VCS theory, and that DA models are calculated with
line profiles that are coherent with the equation of
state, {\it without any free parameter}. In Figure 3, we compare our
improved line profile calculations with the original VCS
profiles. Also shown for comparison are the approximate calculations
of \cite{seaton90}.

\begin{figure}[h]
\centering
\includegraphics[width=30pc]{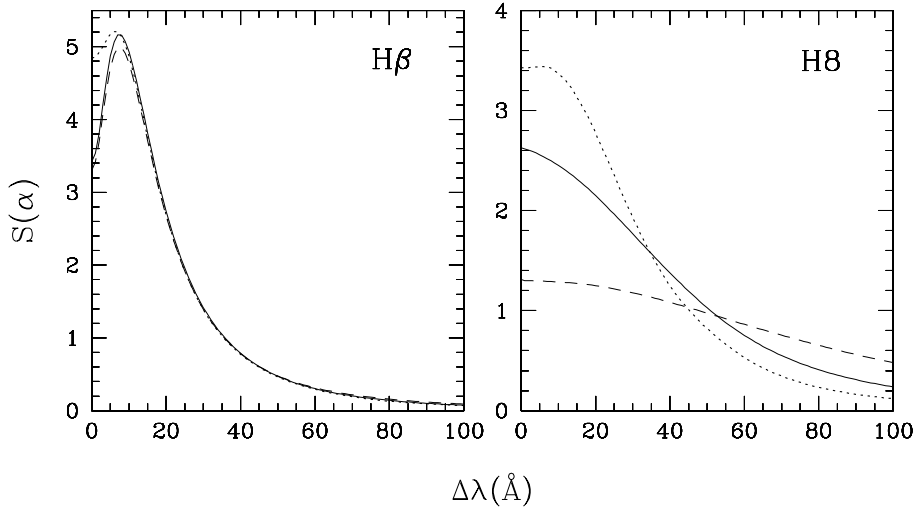}
\caption{Stark broadening profiles as a function of the distance
from the line center (in \AA) at $T=10,000$~K and $\log
N_e=17$. Results are shown for $H\beta$ (left) and $H8$ (right).
In each panel, we compare the results from this
work (solid lines), from the original VCS profiles (dashed lines),
and from the approximate calculations of \cite{seaton90} (dotted lines).}
\end{figure}

\section{Results}

\subsection{Balmer lines}

We now discuss the astrophysical implications of our improved line
profiles. We calculated two grids of model atmospheres using (1) our
new line profile calculations and (2) the original VCS theory using
the tables computed by \cite{lemke97}, but with twice the value of the
critical field ($\beta_{\rm crit}\times2$) as proposed by
\cite{bergeron93}. Both grids range from 13,000 to 40,000 K in 
$T_{\rm eff}$ and from 6.5 to 9.5 in $\log g$. In this analysis, we
chose to restrain our calculations to this range of effective
temperature to avoid the uncertainties related to convective energy
transport, neutral broadening, and non-LTE effects. 

First of all, we can already see in the bottom panel of Figure 1 that
our line profiles have significantly improved the internal consistency
between the solutions obtained from various Balmer lines compared to
the previous calculations displayed in the two upper panels. We find,
however, that the quality of the fits (not shown here) are nearly
identical for all grids. In the following, we analyze the PG
spectroscopic sample of DA white dwarfs from \cite{liebert05}, but we
restrict the range of temperature to that of our model grid. This
yields a sample of 250 objects out of the 348 DA stars for the
complete PG sample. Our spectroscopic data and fitting procedure are
identical to those described in \cite{liebert05}. The atmospheric
parameters for each star are obtained using both model atmosphere
grids, and the $\log g$ values are then converted into masses using
evolutionary models appropriate for white dwarfs with thick hydrogen
layers (see \cite{liebert05} for details). Our results are presented
in Figure 4 in a mass versus $T_{\rm eff}$ diagram. We can see that
the distributions with both grids are significantly different. In
particular, the masses derived with our improved models are globally
larger and the $T_{\rm eff}$ values are also higher (see also
Fig.~1). As seen from Figure 5, the mean mass of this PG sample is
shifted by $\Delta M/M_{\odot} = +0.034$ when our new models are
used. The shape of the mass distribution remains statistically
equivalent for both grids, however, and the dispersions are
comparable. 

\begin{figure}[h]
\centering
\includegraphics[width=25pc]{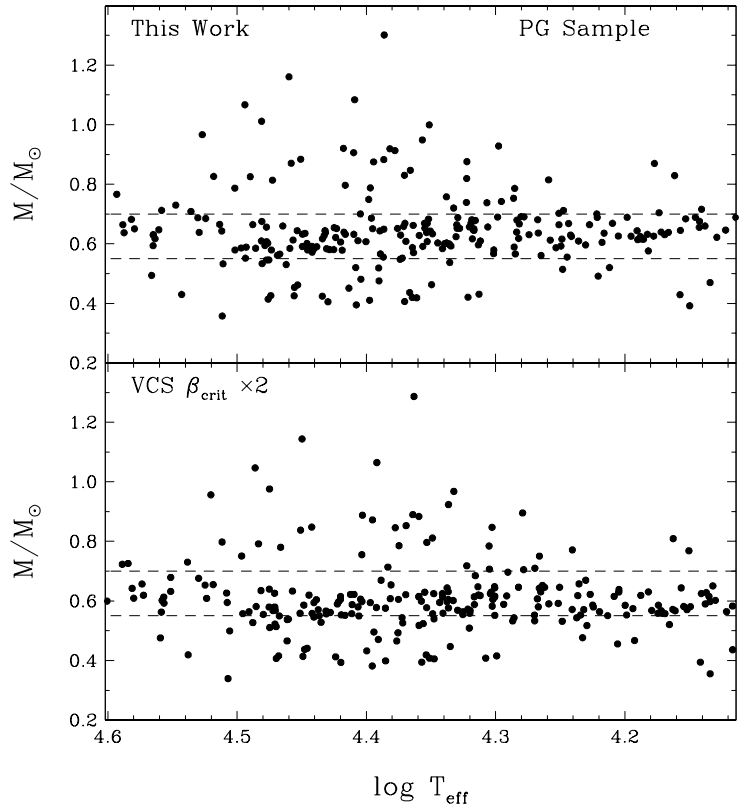}
\caption{Mass versus $T_{\rm eff}$ distributions for the PG sample 
in the range 40,000 K $> T_{\rm eff} >$ 13,000 K. Results are shown
for both our improved line profiles (top panel) and our old models
(bottom panel). Lines of constant masses at 0.55 and 0.70 $M$/$M_{\odot}$
are shown as a guide.}
\end{figure}

\begin{figure}[h]
\includegraphics[width=19pc]{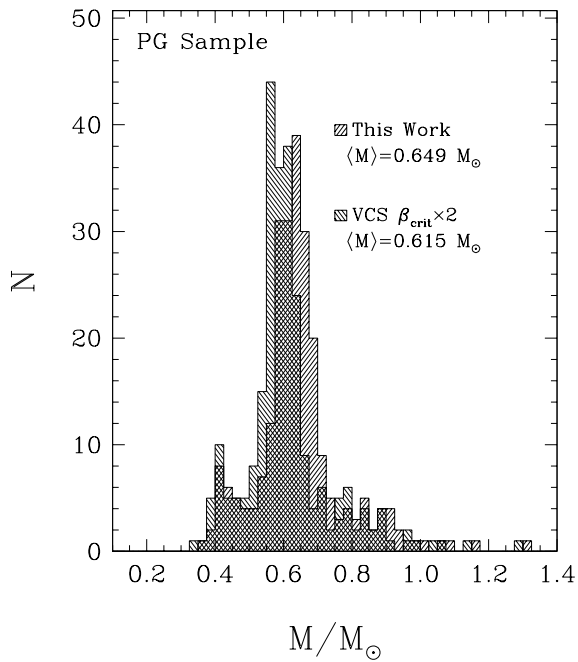}\hspace{2pc}%
\begin{minipage}[b]{17pc}\caption{Mass distributions
for the PG sample used in Figure 4. The mean masses are reported in
the figure.}
\end{minipage}
\end{figure}

As discussed by \cite{bergeron92}, even though the spectroscopic
technique provides very accurate individual determinations of
atmospheric parameters, the {\it absolute} values of log $g$ may
suffer from a zero-point offset and it is therefore important to
compare the individual measurements with results obtained from
independent techniques. The only reliable independent observational
constraints for large samples of white dwarfs are trigonometric
parallax measurements. We thus compare the absolute visual
magnitudes $M_V$ derived from the measured parallaxes (combined
with $V$) with the absolute magnitudes obtained
from the spectroscopic values of $T_{\rm eff}$ and $\log g$ following
the procedure described at length in \cite{HB06}.
The results presented here are based on
the high signal-to-noise spectroscopic sample of
\cite{bergeron07} for 92 DA white dwarfs with known parallaxes. The
comparisons are shown in Figure 6 for both model grids. Again, we have
restrained the analysis to stars above $T_{\rm eff}=13,000$~K. The
agreement is satisfactory within the parallax uncertainties for both
model grids. Hence, despite the fact that our new models yield higher
values of $T_{\rm eff}$ and $\log g$ (i.e., smaller radii), these two
effects almost cancel each other and the predicted luminosities (or
$M_V$) remain unchanged. Another important constraint is
provided by the bright white dwarf 40 Eri B for which a very precise
trigonometric parallax and visual magnitude have been measured by
Hipparcos. These measurements yield $M_V=11.01
\pm 0.01$, while we predict $M_V=11.02 \pm 0.07$ and $10.97
\pm 0.07$ based on our new and old models, respectively. Although both
determinations agree within the uncertainties with the observed value,
our new grid provides an exact match to the measured $M_V$ value.

\begin{figure}[h]
\centering
\includegraphics[width=25pc]{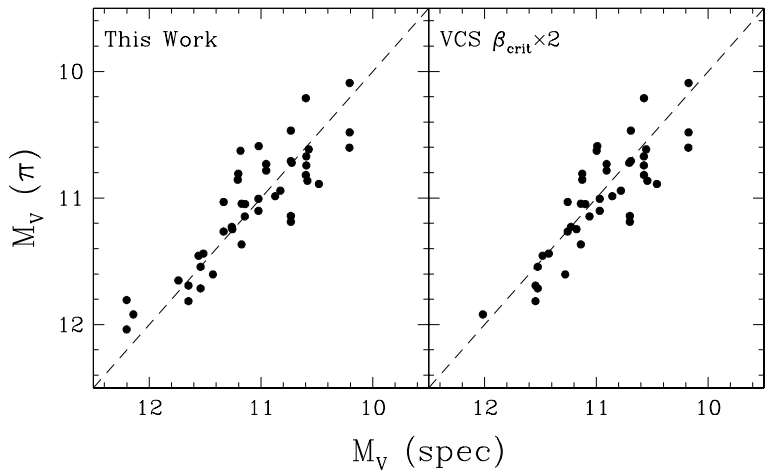}
\caption{Comparison of the absolute visual
magnitudes obtained from trigonometric parallax measurements and from the
spectroscopic analysis of DA stars above $T_{\rm eff}\sim 13,000$~K
using both model grids.}
\end{figure}

\subsection{Lyman lines}

This section presents a preliminary investigation of our model
predictions for the Lyman lines. We analyze the FUSE spectra of two
bright white dwarfs, GD 71 and Sirius B, both of which are in the
temperature range considered in this work. In Figure 7, we show
$\chi ^2$ fits of the entire FUSE spectra obtained with our new models
to determine the atmospheric parameters given in the figure. The
quality of the fits is very good. We performed similar fits with our old
grid and we find that the quality of the fit is comparable, although
our new atmospheric parameters are shifted by +60 K in $T_{\rm eff}$
and by +0.06 in $\log g$ for both stars. Thus, the shifts in $\log g$
are similar to those observed in the optical. It should be noted,
however, that an increase of both $T_{\rm eff}$ and $\log g$ by
$\sim1\%$ results in fits that are equally good from visual
inspection, so the values of the best fit parameters reported here
should be taken with caution. Recent analyses by
\cite{vennes05,barstow03} suggest that the $\log g$ values derived
from the optical and the UV agree in this range of $T_{\rm eff}$, and
this conclusion should not be affected by our new models.

The temperature shifts in the UV, however, are much lower than those
in the optical (see, e.g., Fig.~1). Since the UV spectra were
calibrated using DA models with the older line profiles, the data
reduction might have to be revisited before we confirm these
results. For instance, we show in Figure 8 the predicted spectra over
the visible and UV regions for two flux standards (GD 71 and G191-B2B)
used for the FUSE and HST calibrations among others. The atmospheric
parameters for these LTE models are obtained from the best fits to the
observed Balmer lines of each star using both grids (hotter models
have been calculated for this exercise). The results indicate that the
absolute predicted fluxes, as well as the slopes of the energy
distributions, are very different even up to the high temperature of
G191-B2B ($T_{\rm eff} \sim 60,000$ K).

\begin{figure}[h]
\centering
\includegraphics[width=28pc]{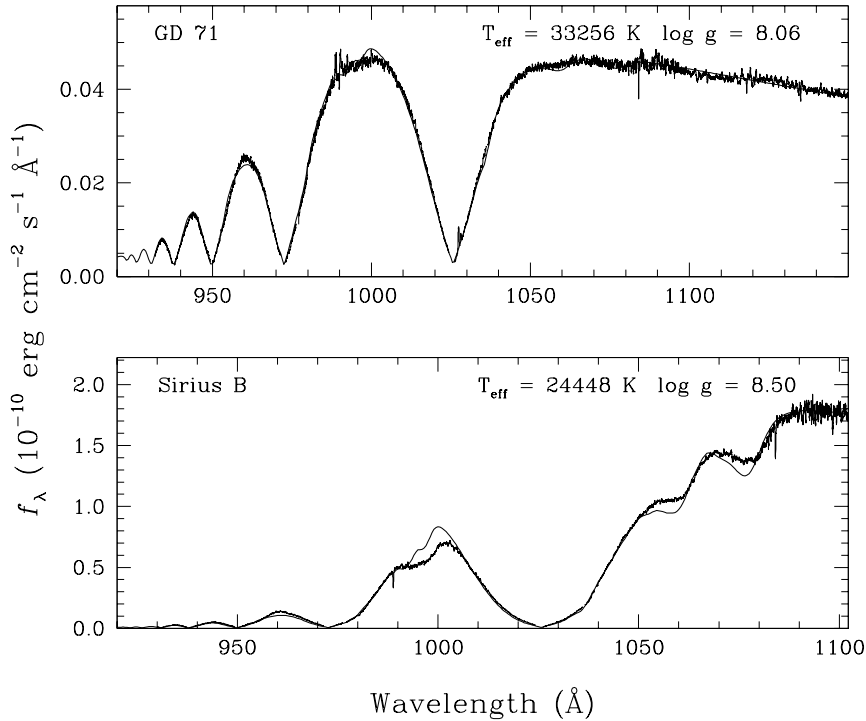}
\caption{Fits to two FUSE spectra using our improved model spectra; $T_{\rm eff}$, 
$\log g$, and the solid angle are considered free
parameters. Interstellar features, including all cores of the lines,
have been removed from the observed spectra. The atmospheric parameters
obtained from the minimization procedure are given in each panel.}
\end{figure}

\begin{figure}[h]
\centering
\includegraphics[width=36pc]{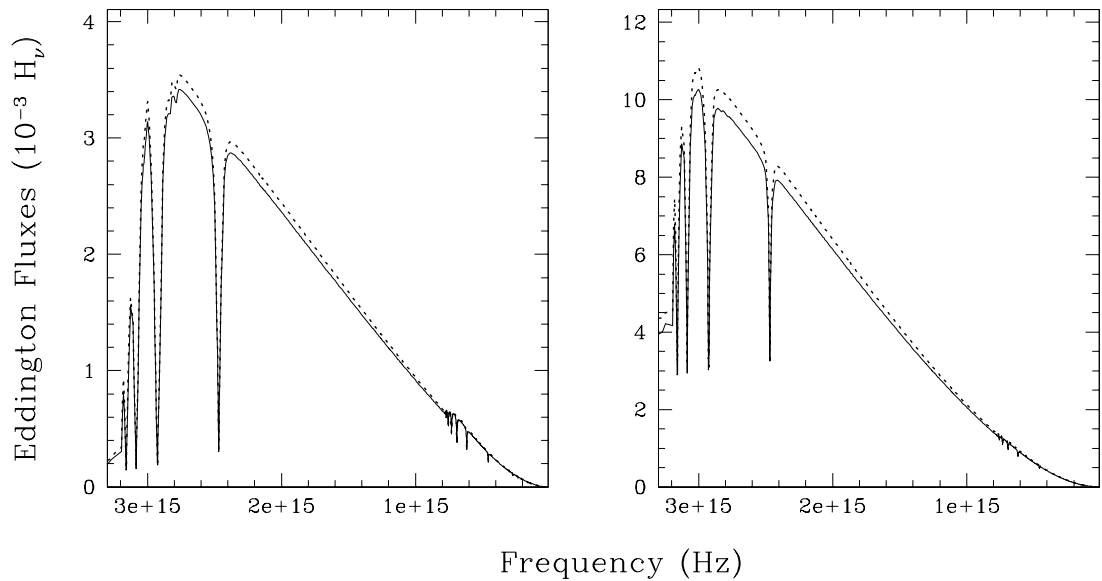}
\caption{Theoretical spectra for the flux standards GD 71 (left) and 
G191-B2B (right) with our improved models (dotted lines) and our old
grid (solid lines) using the atmospheric parameters obtained from fits
to the Balmer lines.}
\end{figure}

We finish this section by noting that the non-ideal effects for the
lower Lyman lines are still not well understood. The critical field in
this case becomes so large that it must have been created by a single
proton close to the absorber in such a way that the approximation of
the dipolar proton-absorber interaction fails completely. For
instance, exact calculations show that $n=1, 2$ and $3$ have no
crossings. Current white dwarf model atmospheres (this code, TLUSTY)
neglect non-ideal effects for Lyman $\alpha$, Lyman $\beta$ and in
some cases Lyman $\gamma$. In other words, there is no
pseudo-continuum opacity, and ideal gas profiles are used for these
lines. This significantly changes the predicted UV spectra for $T_{\rm
eff}<30,000$~K. For H$\alpha$, non-ideal effects are already
negligible with the HM88 theory and no modification is necessary. We
plan to study further this problem with a larger sample of FUSE
spectra.

\section{Conclusion}

We have combined for the first time the unified theory of Stark
broadening from VCS with the non-ideal equation of state of HM88 to
build model atmospheres for DA white dwarfs that are consistent to
first order in terms of the physics of the equation of state and the
line profiles, without a free parameter. We have shown that in the
range 40,000~K $ > T_{\rm eff} >$ 13,000~K, the revised mean mass for
the PG sample is significantly larger than previous
determinations. The atmospheric parameters we obtain with our new
model spectra are in excellent agreement with trigonometric
parallax measurements. Future work will extend our analysis to the
more complex regime of white dwarf stars at higher and lower effective
temperatures.

\ack{We thank M.~Lemke for providing us with his computer version of the VCS code. This
work was supported in part by the NSERC Canada and by the Fund FQRNT
(Qu\'ebec). P. Bergeron is a Cottrell Scholar of Research
Corporation.}

\end{document}